\documentstyle{npbproc}

\begin{document}
\title{NON--PERTURBATIVE BOUNDS ON THE HIGGS MASS IN THE MINIMAL STANDARD MODEL
\thanks{This is a slightly extended version of the talk delivered  at the
Topical Workshop ``Non perturbative aspects of chiral gauge theories'',
Accademia Nazionale dei Lincei, Roma, 9-11 March, 1992. }
}

\author{Herbert Neuberger}
\address{Department of Physics and Astronomy \\
          Rutgers University,  Piscataway, NJ 08855-0849, USA}

\date{}

\runtitle{Non--perturbative bounds on the Higgs mass in the minimal standard
model}
\runauthor{H. Neuberger}

\pubyear{1992}
\volume{XXX}
\firstpage{1}
\lastpage{14}

\begin{abstract}
The Higgs mass in the minimal standard model is bounded by triviality and
vacuum
stability in the range 50--100 $GeV$ to 700--900 $GeV$. Recent results will be
presented in brief and directions for future work will be proposed.

\end{abstract}

\maketitle

\section{INTRODUCTION}
Chiral gauge theories are really pretty when you don't have scalar fields
around. Unfortunately
I'll be talking only about the ugly stuff, and even have the gauge degrees of
freedom turned off
altogether.

The experimental verification of the standard model has become so successful
because we were able
to explore energy regimes in which the forces are weak, and there measurements
of scattering processes
identify the interaction Hamiltonian directly by the Born approximation. We
have had only quite
limited success in controlling  the strong color force quantitatively, but we
were able to convince
ourselves that the problem had been ``contained'', and moved ahead.

Only two couplings are left that could be strong at contemplated accelerator
energies: the Higgs
selfcoupling and the top quark Yukawa coupling. About the Higgs particle we
know little from
experiment; the top quark however seems to be on the verge of discovery and
there are indications
that its Yukawa coupling, although much stronger than any other Yukawa
couplings, is still quite weak.

The minimal standard model is  a set of approximate calculational rules, and we
have evidence
that the approximation must break down if the Higgs mass is heavier than 800
$GeV$ or so\cite{Einhbook}. If the top
is in the range 100---200 $GeV$ there is a calculation that shows that the
Higgs can't be lighter
than 50---100 $GeV$ approximately\cite{Sherrev}. Can we state therefore,
excluding ``naturalness'' preconceptions,
that we already know that $m_H \simeq 450 \pm 350~ GeV$ if a Higgs particle of
mass $m_H$ at all exists?
\footnote{This ought to be compared with some older
and some more recent
estimates of limits on the Higgs mass obtained by
comparing radiative
corrections with high precision experiments:
Ellis and Fogli\cite{EllFog} obtained
$1.8~GeV < m_H < 6000~GeV$ at 68\%
CL in 1990, and updated the numbers
(in collaboration with Lisi\cite{EllFogLis})
to $1.4~GeV < m_H < 160~GeV$ at 68\% CL and
$0.5~GeV < m_H < 1500~GeV$ at 90\% CL.
The present most preferred values are, according to this
last analysis, $m_t \simeq 120~GeV$ and $m_H \simeq 10~GeV$.}
I think that with a few more numerical checks  and analytical
computations we shall be
able to responsibly make a statement of this kind within a year or two.
It would be nice if the
bounds came out tighter at the end.

\section{ UPPER BOUND}

Let me briefly review the ``trivial'' logic\cite{neubcapri}. Assume that the
minimal standard model is embedded in some other theory, whose effects become
apparent at a scale $\Lambda$ that is larger than $m_H$ and $f$ by a sufficient
amount to make
an expansion in ${{\mu}\over{\Lambda}}$ a good approximation for energies $\mu$
lower than two to three times
$m_H$. Assume also that ${{m_H}\over{f}}$ is quite large, so that one can
ignore even the heaviest fermion. These assumptions mean that there exists an
effective action (in the sense described by
Georgi\cite{Georgibook}\cite{Georgitalk}) which, in the purely scalar
sector, including all {\sl observable} leading effects of the cutoff, is
parameterizable by four numbers (after taking into account the freedom of
field redefinitions). If ${{m_H}\over{f}}$ is not too large one can use
perturbation theory in the effective theory; the very definition of the
effective theory is then set in the perturbative expansion. If the scalar
selfcoupling is too large, it is impractical to use the effective
theory. We assume however that this effective theory does exist, even outside
perturbation theory. It's mere existence implies, by running the arguments
that are used when the effective theory is constructed from its underlying
model
in the reverse, that there exist many lattice actions on the $F_4$ lattice that
{\sl exactly} reproduce the same physical observables to this order in the
inverse cutoff. Any one of these actions would lead to identical effective
theories if we declare that we are uninterested in contributions of order
$1/\Lambda^4$ multiplied by any power of $~\log (\Lambda )$. Such effective
theories
are formally represented by ``Symanzik Local Effective
Lagrangians\cite{Sylel}''. Again,
when the renormalizable coupling is strong, the computational value of the
effective action is limited. However, we know now that if we could solve
all $F_4$ lattice scalar field theories, we would find in the list of
solutions also an exact representative of the embedded standard model,
including
order $1/\Lambda^2$ effects. All this is useful only as long as an expansion
in $1/\Lambda$ is, numerically, a good approximation at leading order. If it
is not, the effective action contains an infinite number of important
parameters, and reproducing it in its entirety by some scalar $F_4$ lattice
field theory is an unachievable goal in practice. Because of this we made the
additional assumption stated at the beginning of this paragraph. I think that,
in the absence of this assumption, one cannot make meaningful statements
regarding the ``triviality bound''\footnote{
This is not a real limitation, when the assumption is violated it simply means
that the
symmetry breaking mechanism cannot be represented
by renormalizable scalar field interactions with any reasonable accuracy. }.
The practical importance of the assumption
is that it limits the number of unknown parameters in the lattice action
that have to be varied. This turns the problem of estimating the physically
relevant triviality bound into a finite problem, that, fortunately, can be
reliably solved with present day computer resources. In this sense, the upper
bound on the Higgs mass is an almost unique physically relevant nonperturbative
problem.

Suppose we want
to represent on the $F_4$ lattice as heavy a Higgs particle as possible. We
start by making some educated guesses for what kind of bare lattice actions
we should look at:
We can say with reasonable certainty that if we use a four component real field
for the Higgs doublet on the lattice, the maximal
value for the Higgs mass will be obtained when the bare four point coupling in
the lattice Lagrangian becomes infinite.
Therefore, the upper triviality bound will be obtained in a bare non--linear
model. Out of four parameters we are left with only three. These three
parameters can be thought of as the coefficients of the terms of
second order and fourth order in derivatives in the bare, non--linear, lattice
action.
The coefficient of the leading term simply sets the lattice spacing in terms
of $f$, i.e. tells us how fine the lattice is in $GeV$ units. Tuning it, we are
varying the cutoff $\Lambda$. All work until now simply set the additional two
coefficients to some arbitrary fixed values and tested how low can one make
$\Lambda$
within our assumptions.
This led to the finding, as is well known by now, that simplest lattice actions
allow Higgs masses larger than $\sim
600~GeV$\cite{LW3}\cite{HCMC}\cite{BBHNnum} only for quite small inverse
lattice spacings, of the order of $1~TeV$ on the $F_4$ lattice.\footnote{ The
corresponding inverse lattice spacing on the hypercubic lattice is larger, but
this is illusory, because Lorentz violation effects are much stronger, as
can be verified from tree level perturbation theory; a hypercubic lattice with
inverse nearest neighbor distance 2 $TeV$ is coarser physically than an $F_4$
lattice whose inverse nearest neighbor distance is also 2 $TeV$\cite{BBHNana}.}

It may seem then, that a strongly interacting Higgs sector without
a host of other cutoff effects is ruled out. However, if one recognizes that
the real content of the results is that ${{m_H}\over{f}}$ cannot be larger than
$\sim 2.4$, once $m_H < \Lambda$, one may be surprised and worried that the
action chosen happened not to be general enough\cite{neubtalla}. For example, a
QCD like theory
is commonly assumed to have ${{m_H}\over{f}}~\sim 6$
albeit with $m_H \sim \Lambda$. There is a lot of ``room'' between
$2.4$ and $6$ hinting that the simplest action may be a bad place to estimate
the triviality bound from.  A systematic search in the parameter space $p$ of
other lattice actions is needed before a more general statement can be made.
Moreover, large $N$\footnote{
$N$ is the number of real components of the scalar field in a model where
$SU(2)_{weak}  \times SU(2)_{custodial} \approx O(4)$ has been promoted to
$O(N)$.
} calculations done by Einhorn\cite{EinhN}, have indicated that cutoff
models with a reasonably strongly interacting Higgs particle do exist. While
this indication holds only for $N$ large and
it is known that there are large finite $N$ effects at the physical
value of $N=4$\cite{Dashneub}\cite{Lin}, the existence of strong interactions
in scalar field theory
could be viewed as a {\sl qualitative} rather than a quantitative property.
The results obtained with the simplest action on the hypercubic lattice
for the upper bound on the selfcoupling in the broken phase have shown a
remarkable qualitative and quantitative smoothness when going from $N=1$ to
$N=4$\cite{HCMC}\cite{BBHNnum}\cite{LW12}\cite{Kuti1}, and it is obvious that
there does exist a qualitative difference between $N=1$ and $N\ge 2$, because
the $N=1$ model has no Goldstone bosons. It is
hard to see why the situation at higher values of $N$ would change
qualitatively. Einhorn implicitly employed a sharp momentum cutoff in Euclidean
space
in his $N=\infty$ work;  this cutoff is not acceptable because it
does not lead to cutoff effects that can be parameterized by an effective
action
of the usual type\cite{neubcapri}, but this problem can easily be fixed.
We cannot ignore results obtained in the $1/N$
expansion simply because we are interested only in $N=4$.
Rather, we are justified to suspect that the reason that only weakly
interacting
Higgses have been produced on the lattice until now has more to do with the
type of lattice actions that were used than with the value of $N$\cite{HNV}.
Thus,
it makes sense to employ the $1/N$ expansion also in our search for the
appropriate lattice action. I shall say more about $1/N$ later.

Let me review some recent results obtained on this problem by Heller, Klomfass,
Vranas  (from SCRI and FSU) and myself\cite{HNV}\cite{HKNV}. Define a function
$C(m_H /f  ; p)$:
$$
C(m_H  /f ; p)=
$$
$$
{{m_H}\over{\Lambda}}  \lbrack {{(N+8)m_H^2}\over{(4\pi f)^2}}
\rbrack^{{9N+42}\over{(N+8)^2}}
\exp \lbrack {{(4\pi f)^2}\over{(N+8)m_H^2}}\rbrack
$$
Here $p$ denotes all the parameters of  the bare action, except one that turns
into the relevant direction in the immediate vicinity of  the massless theory,
which has
been traded for $f$. More specifically,  one works on a one dimensional line in
the space of bare couplings; the line connects some point in the broken phase
to a point on the critical surface. The line is defined by keeping the
parameters $p$ fixed and $f$ can be used as a coordinate along the line. Going
along the line, away from the critical surface, one finds a bound at the point
where cutoff effects become too large. Varying the set of parameters $p$ one
than can maximize over all  ``line bounds''.

The function $C(m_H /f  ; p)$ admits a power series expansion in $m_H /f$. The
evaluation of $C=C(0;p)$ is a strong coupling problem; $C$ depends on the
physics at cutoff scales. To prevent Lorentz breaking lattice artifacts from
mixing into genuinely realizable ``new physics''  effects at order
$1/\Lambda^2$  we  concentrated in our numerical work only  on the $F_4$
lattice\cite{neubf4}.  Our
problem is to find the region in $p$ space where $C$ is as small as possible;
a small $C$ means that larger values of  $m_H / f$ do not require too large
matching increases in the
prefactor $m_H / \Lambda$ which signal the set in of uncontrollable cutoff
effects.

In numerical simulations one sets $N=4$ because this is the correct value in
the
standard model. Analytical studies, however, can be performed with equal ease
at arbitrary values
of $N \ge 2$. In particular, one can employ the $1/N$ expansion to learn many
semi--quantitative
things about the system. Direct applications of $1/N$ are not reliable
numerically, because, as is clear from the
equation, the ``real'' expansion parameter is $1/(N+8)$ to one loop order and
this is very different from $1/N$ when $N=4$\cite{Dashneub}\cite{Lin}. It is
possible that one doesn't have to include more than one or two more orders in
the $1/N$ expansion and then, with the help of Pad\' e resummation, one may get
sufficiently high accuracies to make $1/N$ quantitatively reliable. In practice
it was found however, that, even the leading order term, gives much more
accurate predictions when the
coupling $m_H/f$ is relatively large, and the one loop term doesn't dominate
any more. This can be tested by comparing numerical simulations at $N=4$ with
infinite $N$ formulae for the same lattice
action. Typical orders of magnitude of deviations between infinite $N$ and
numerical results are 20\% --30\% for the relation between $m_H /\Lambda$
and $m_H /f$ in the vicinity of the bound. It makes sense therefore,
when testing the dependence on the parameters $p$, to first do a large
$N$ analysis to locate regions in parameter space where a higher bound
is likely and only subsequently
go to the computer. The gross features of the relevant
region in  the space of actions at infinite $N$ are schematically
represented in the figure below:
\vskip 3.4in

We showed first that at infinite $N$ one out of the two additional
parameters has no effect on the bound, while the other affects the Higgs mass
strongly. We also understood this result in physical terms, so that it appears
reasonable to
expect that it survives at $N=4$ too. We tested our physics intuition by
working, at infinite $N$, with
different regularization schemes, not only on the lattice, and concluded that
our picture is indeed not an artifact of one particular regularization scheme.

Let me first briefly recall some older results, corresponding, roughly, to the
line $Z=1$ in the
schematic phase diagram: With the simplest bare action Bhanot, Bitar, Heller
and I\cite{BBHNnum}
had obtained earlier  $C = 18_{-5}^{+8}$.  More detailed old numerical results
are summarized
in the  graph below:
\vskip 3.4in

The large errors in $C$ are  acceptable because the exponent transforms them
into small
errors in $m_H/f $. For larger  $m_H/f$ the asymptotic formula doesn't work as
well and,
working on lattices of sizes up to $14^4$ we obtained, for example,  $m_H/f
\simeq 2.4$ at  $m_H/\Lambda=.7$ and  $m_H/f \simeq 2.0$ at $m_H/\Lambda=0.3$.

Using more complicated nonlinear actions, which can still be chosen to be {\it
nearest neighbor only},  we have shown that, in the large $N$ limit, $C$ can be
decreased by a factor of 4---5 when the relevant four derivative coupling is
turned on to maximal acceptable strength,
and this induces an increase in the upper bound by 25-30\%.
Recently we obtained our first results from computer work at $N=4$, on the
$F_4$ lattice.
We found, in the  physical case $N=4$, that $m_H/f \simeq 2.4$ at
$m_H/\Lambda=.3$
which is an increase by 25\%, in  reasonable agreement with the calculations.
Phrased differently, a Higgs mass of about $600~GeV$ can be realized by the
more
complicated action with scaling violations that are smaller by a factor
of about 5.4 ($({{.7}\over{.3}})^2$) than the cutoff effects that necessarily
accompany
a similar physical Higgs mass when the simplest nearest neighbor action is
used.\footnote{
At $N=\infty$ the $\pi$--$\pi$ propagator has zero physical (on shell) scaling
violations at order $1/\Lambda^2$ independently of the additional parameters
$p$; the physical
quantity most naturally related to the propagator is whatever plays the role of
the Lehmann spectral
density. We also confirmed explicitly that on the $F_4$
lattice minimizing $C$ over $p$ reduces the cutoff effects at
fixed physical Higgs mass in other physical observables.}

The heaviest Higgs that we have produced to date has a mass of $\simeq 700
{}~GeV$.
We have difficulties going above this because finite width
effects cannot be ignored any more. On the $F_4$ lattice as long as the
Higgs mass is smaller than ${{2\pi \sqrt{3}}\over {L}}$ the decay
cannot occur because even the lowest energy pion pair is too heavy. $L$ cannot
be made too small because the Goldstone dynamics
must be still reproduced on the finite lattice.
For typical $L$ values $(\sim 16)$, we are safe if $a_{F_4} m_H \le .6$. A
lattice Higgs mass of $.6$ corresponds to a physical mass $m_H \simeq 700~GeV$.

With the increase in mass the width increases and the systematic error
associated with the finite width ($\sim \Gamma /2$) becomes of the order of
12\% and grows rapidly, to 16\%, at a mass of $\sim 800 ~GeV$. That
finite width effects cannot be ignored any more is indicated by the departure
of the RG predictions
(augmented by tree level and one loop scaling violations) from the measured
masses at the last points in the graph obtained in the simplest nearest
neighbor analysis. A {\sl preliminary} set of data resulting from the new
SCRI/FSU-Rutgers collaboration\cite{HKNV} is presented in the graph below:
\vskip 3.4in

Our main lesson to date is a  ``thumb rule'' for making $C$ smaller,
that is  the bound larger: Write a bare action that introduces as much
repulsion between two pions in a $I=J=0$ state as is possible without
altering the vacuum  structure of the  of the system. This delays in
energy as much as possible the formation of the Higgs particle as an
unstable bound state in the $I=J=0$ channel. Actions that are close to
saturating
the bound have a maximal momentum ($\pi/a$ where $a$ is the lattice spacing)
$\Lambda \approx 1.5 \pi m_H \approx 4\pi f_\pi $ (since $m_H / f_\pi \approx
2.6$). This is the most ``natural'' relationship between $\Lambda$ and $f_\pi$
in a chiral
Lagrangian as explained in Georgi's book. It is therefore reasonable to use
``chiral
Lagrangian logic'' when trying to understand the physical effects generated by
such
actions.

Our result shows that the previous bound was too low, but,
also, that it is quite hard to make the Higgs heavier, and even a serious
effort, could not yield more than
a $30$\% increase above $m_H \simeq 600~GeV$. Therefore, although it
appears now that we cannot say
anymore that a strongly interacting Higgs has been ruled out,
since a Higgs of $\sim 800~GeV$ would have a width
of  $\sim 250~GeV$ and would selfinteract strongly by any
reasonable standards, we still can say that it is
perfectly reasonable, when trying to make predictions within the
standard model, not to allow the unknown
Higgs mass parameter to exceed 800--900 $GeV$.

When can we stop in our search  for the bound in the space of actions? The
bound is quantitatively
meaningful only if the cutoff effects are still small and hence parameterizable
by a few parameters.
The ``thumb rule'' provides a further restriction in this space: with its help
I believe that the
search has been sufficiently narrowed down that with a few more checks and
maybe some new
input from simulations  on hypercubic lattices\footnote{
Recent work by F. Zimmermann\cite{zimmer} indicates that the bound will go up
on hypercubic lattices too.}
we shall have, essentially
for the first time, an estimation for the triviality upper bound that
deserves to be taken seriously by the rest of
particle physicists in the sense that it cannot be viewed any more as just a
property of some lattice
regularization, but is claimed to be a property of $\sl any$ embedding of the
minimal standard model
into a theory with an energy scale significantly higher than $1~TeV$.

For the more distant future I would propose to specifically
work on disproving or proving the following statement (or a somewhat altered
form of it):

{\sl  $m_H > 800$--$900~GeV$ will be possible only if the next lightest
resonance in pion--pion
scattering is a $\rho$--like particle (with $I=J=1$).  Increases in $m_H /f$
above 3.5 can be obtained
only when accompanied by decreases in $m_\rho / m_H $ to somewhere between 1
and 2.}

This statement is true if the single solution to the unitarity, symmetry and
low energy constraints
on the S-matrix of pions interacting with a Higgs particle heavier than a given
number (3--4) times
$f_\pi$, is the one found by a QCD--like theory.\footnote{
I have no compelling argument supporting the above statement; I made it because
I think that it could provide a useful focus for the next year or so. I do
think that the statement has a reasonable chance to be true because approximate
enforcement of crossing, unitarity and soft pion theorems have been shown to be
quite restrictive under some assumptions\cite{brown}.}
This would mean that to make the Higgs much heavier
one needs to postulate something like technicolor. But the latter might have
been already
experimentally excluded by the sign of the
$S$ parameter\cite{spar}. This exclusion is independent of fermion mass
generation
mechanisms (technicolor extension),
and
essentially, might follow directly
from the existence of the techni--$\rho$ like particle. What I mean is that the
$S$ parameter
can be determined directly from the structure of the scalar sector, without any
need for a
technicolor detour; we used technicolor only as an indication that, similarly
to that case,
numerical estimates of the $S$ parameter may turn out to be quite restrictive.
Since the $S$
parameter measures the difference between two $U(1)$ currents at small momenta
(one of
the $U(1)$'s is spontaneously broken, while the other is preserved), it is
quite likely that $S$ can be
extracted from a finite size relation somewhat similar to the one used to
obtain $f$ in the
scalar $O(4)$ simulations\cite{neubfpiprl}\cite{hellerneub}\cite{leut}.

In summary, I think the following steps would be worthwhile to take in future
lattice investigations of a more or less strongly coupled scalar sector:

\noindent
(1) Develop good numerical ways to measure $m_\rho$.\footnote{
Very recently, after the conference, Bitar and Vranas\cite{BitVra} circulated a
preprint with evidence
that no $\rho$ particle with a reasonably small width is being produced
by the simplest hypercubic lattice action. These authors came up with an
effective method for extracting the $\rho$ signal from the numerical data.}

\noindent
(2) Analytically estimate (within $1/N$) the $\rho$ mass. The analytical
calculation is a
bit messy because one must go to subleading order in $1/N$ in pion-pion
scattering; at
leading order the
amplitude is dominated by scalar exchange and no fixed poles in Mandelstam's
$t$ variable occur.

\noindent
(3) Start including the $S$ parameter in numerical and analytical ``triviality
work''.
There is an interesting possibility that a combination of high precision
experimental
data in tandem with triviality considerations will tighten the upper bound.

\noindent
(4) Account for finite width effects in numerical simulations. This should
be possible because the number of energetically accessible two pion states
is very small for realistic volumes (in most case just one state is below
threshold).

\noindent
(5) The computation of scaling violation effects (effects of
higher dimensional operators) on the lattice still suffers from the absence of
a good resummation procedure for the leading logarithms in even the first
term. Again $1/N$ has the potential to help. It may be more reliable
numerically than simply truncating the loop expansion (which is what is done at
present\cite{LW3}\cite{BBHNana}\cite{LW12}).

\section{LOWER BOUND}

Let me start by briefly reviewing what we know from experiment. The combined
LEP data on direct searches for the Higgs in $e^+ e^- \rightarrow H~Z^*_
{ \hookrightarrow f \bar f  }$ yield a lower bound on $m_H$ of $57 ~GeV$. LEP
will be able to get to $\sim 60~ GeV$; LEP 200 can explore up to $\sim90~
GeV$ with a real, rather than virtual, Z in the final state. LEP won't be able
to distinguish between a
minimal standard model light Higgs and a light Higgs of a supersymmetric
extension of the standard model. The mass of the top is known to be larger
than $89~GeV$ (CDF) and radiative corrections bracket it approximately
by $120 ~GeV \le m_t \le 180 ~GeV $\cite{phen}.

The latest theoretical estimate coming from ``vacuum stability''  that I am
aware of\cite{SherZ} is a lower
bound on the Higgs mass
that increases about
linearly between $\sim 30 ~GeV$ and $\sim 50~ GeV$ when $m_t$ increases in the
above interval
and the effective cutoff of the standard model is taken at $\Lambda = 10^3
{}~TeV$.
For $\Lambda = ~10^4 ~TeV$ the range is $\sim 40~GeV$---~$\sim 100~GeV$ and for
$\lambda=10^{15} ~TeV$ it is $\sim 45 ~GeV$---$\sim 160~ GeV$. I haven't
seen answers quoted for smaller values of $\Lambda$.
In view of the experimental situation it is likely that he triviality
upper bound will outlive the stability lower bound.

The numerical value of the lower bound is sensitive to the color  gauge
coupling
and it is hopeless to attack the full problem directly by the lattice methods
of today.
All that one can do is to test the
validity of the perturbative calculational scheme in a simpler model. However,
I am unaware of any
solid argument that would throw serious doubt on the perturbative analysis,
mainly because the top, while heavy, appears to have a Yukawa coupling well
below half its tree level unitarity bound.

There are several lattice studies that intentionally or indirectly have some
bearing
on the lower bound\cite{revYuk}. It is mainly because of the availability of
some numerical data that
I'll spend some time on this topic:

Following Krive and Linde\cite{KrLd} consider the simplest prototype model, a
real single
component scalar field coupled to a massless Dirac fermion by a Yukawa
coupling. There are two slightly
different approaches to the bound issue. One is based on the RG flow of the
couplings and originated
with Cabibbo, Maiani,  Parisi and Petronzio\cite{Cabetal}; the other
used by Krive and Linde originally, is based on the
evaluation of the effective potential $V_{eff} (\phi_c^2 ) $, where $\phi_c$ is
the ``classical'' scalar field.

Let me first review the RG approach: Introduce the logarithmic energy scale
$t=\log {\mu \over f}
$ where $\mu$ is some energy and $f$ is the physical scale defined more
precisely below. Just by
inspection of the relevant one loop diagrams one sees that the 1--loop RG
equations will have the
following structure:
$$
{{d\lambda }\over{dt}} = b_1 \lambda^2 +b_2 \lambda g^2 -b_3 g^4
$$
$$
{{dg}\over{dt}}=b_4 g^3
$$
where $\lambda$ is the scalar selfcoupling, g is the Yukawa selfcoupling and
$b_{1,2,3,4} >0$. Define $\alpha(t) ={{\lambda(t)}\over{g^2 (t)}}$ ; with
proper normalizations
$\alpha (0) = {{m_s^2 }\over {m_f^2}}$ where $m_s$ and $m_f$ are the scalar
and fermion mass respectively. The RG equations imply
$$
{1\over{g^2 (t)}} {{d\alpha}\over{dt}} = b_1 \alpha^2 +(b_2 -2b_4 )\alpha -b_3
$$
$$
\equiv b_1 (\alpha
-\alpha_{-} )(\alpha - \alpha_{+})
$$
with $\alpha_{+} \alpha_{-}=-{{b_3}\over{b_1}} < 0,~\alpha_{+}> 0$. The
equation for
$g$ gives ${1\over{g^2}} =
-2b_4 (t-t_L )$ where $t_L=\log ({{\mu_L}\over{f}})$ and $\mu_L =f \exp
(1/(2b_4
g^2(0)))$ is the Landau pole energy associated with $g(0)$. Define $\tau = \log
(t_L -t)=\log(\log({{\mu_L}\over{\mu}}))$; then the equation for $\alpha$
becomes:
$$
-{{d\alpha}\over{d\tau}} ={{b_1}\over{2b_4}} (\alpha
-\alpha_{-})(\alpha-\alpha_{+})
\equiv \rho (\alpha )
$$

$\rho (\alpha)$ is a parabola; for $\tau$ decreasing, the flows between the
roots
will go towards negative $\alpha $ values and outside the interval between the
roots it will go towards positive $\alpha$ values. $\tau$ decreases when $t$
increases; we shall always assume that $t$ is smaller than $t_L$. $\alpha_{+}$
is seen therefore to be an infrared fixed point: if we start from some large
$t$
with some reasonable value (order 1) we shall end up at $\alpha_{+}$ when
$t=0$.

However, if ${{\mu_L}\over{\Lambda}} \gg 1$ where $\Lambda$ is the real cutoff
of
the theory, and $1\ll{{\Lambda}\over{\mu}}\ll{{\mu_L}\over{\Lambda}}$, $\tau$
changes
very little between $\Lambda$ and $\mu$ (${{d\tau}\over{d\log
\mu}}=-{{1}\over{\log
({{\mu_L}\over{\mu}})}}$) and hence $\alpha$ can be safely
viewed as a free parameter, allowing arbitrary scalar and fermion masses
in the perturbative domain, as usual.

Suppose now that $0<\alpha (t=0) \ll\alpha_+$; if $\alpha (0)$ is sufficiently
close to zero even the slow variation of $\tau$ with energy will, eventually,
make $\alpha$ negative at the cutoff energy $\mu =\Lambda$. More precisely,
with
$t_L = {1\over{2b_4 g^2 (0)}}$, for $t \ll t_L$, $\tau\approx \log(t_L )-
{t\over{t_L}}$, we get $\Delta \tau = -2b_4 g^2 (0) \Delta t$. Therefore,
as long as $\Delta \tau\equiv -2b_4 g^2 (0) \Delta (\log\mu
)\ge\alpha(t=0)\equiv{{\lambda (0)}
\over{g^2 (0)}}$ we shall hit negative values of $\alpha$ before we hit the
cutoff. However,
if $g^2 (0) \ll 1$, a large variation between the physical scale $f$ and the
cutoff
$\Lambda$ can be accepted without destabilizing the vacuum. In summary,
depending on the assumed value of $f/\Lambda$, we get a lower bound on $\alpha
(0)$
which in turn implies a lower bound on the scalar mass if the fermion mass
is taken as given. If $f/\Lambda$ is let to be close to one, the bound becomes
very weak, while if $f/\Lambda$ is small (but not smaller than $f/\mu_L$), the
bound on the mass ratio approaches $\alpha_+$ from below.

In the effective potential approach of Krive and Linde, one calculates
$V_{eff}$
in renormalized perturbation theory with assumptions violating the above bound:
i.e.
$\Lambda \gg f$, $\alpha (0) \ll1 $, and ${{\lambda (0)}\over{g^4 (0)}}\sim1$.
This can be arranged with
$g^2 (0) \ll 1$ and $\lambda (0) \ll g^2 (0)$ and therefore validate
the use of the loop expansion. For not too large values of
$\phi_c^2$, $V_{eff}$ is dominated by the scalar tree level contribution ($O(
\lambda )$), and by the one fermion loop contribution, relatively of order $O(
(g^4
/\lambda )\log (\phi_c^2 /f^2 ))$. The one loop scalar contribution of
relative order ($O(\lambda \log (\phi_c^2/f^2 ) )$) and the two and higher loop
fermion contributions of relative order ($O(g^2 \log (\phi_c^2 /f^2 ))^k$) are
negligible if the field is not made too large. Defining $f^2$ to be the
location
of the first minimum of the effective potential we obtain, with $\sigma =
{{\phi_c^2}
\over{f^2}}$,
$$
f^{-4} V_{eff} (\phi_c^2 ) =
$$
$$
{{\lambda}\over{2}} (\sigma -1)^2 - b_5 g^4
\int_1^{\sigma} dx x \log x =
$$
$$
{{\lambda}\over{2}} (\sigma -1)^2 - {1
\over 2} b_5 g^4 [\sigma^2 (\log \sigma ~ - {1\over 2}) + {1\over 2}]
$$
where $b_5 >0$ is some number.\footnote{The $g^4$ term simply represents
the sum of all the shifts towards lower energies of the states that make up
the Dirac sea:
$$
\Delta E = -2 \int_{\Lambda} {{d^3 k}\over{(2\pi )^3}} \left ( \sqrt { k^2 +
g^2 \phi_c^2 }-|k|\right )
$$
The factor 2 counts spin states, $g\phi_c$ is the fermion Dirac mass, the
integral is regulated by some cutoff scheme dependent on $\Lambda$ and $\Delta
E$
is the change in the vacuum energy density. For small
${{g^2 \phi_c^2}\over{\Lambda^2}}$ one can expand:
${{\Delta E}\over{\Lambda^4}}\approx a{{g^2 \phi_c^2}\over{\Lambda^2}}+b{{g^4
\phi_c^4}\over{\Lambda^4}}
(\log{{g^2 \phi_c^2}\over{\Lambda^2}} +c) +O({{g^6 \phi_c^6}\over{\Lambda^6}}
\log{{g^2 \phi_c^2}\over{\Lambda^2}})$. The form displayed in the equation in
the text
is now easily obtained by adjusting the tree level contribution to $V_{eff}$
appropriately.}
It is easy to see
that it doesn't take a too large $\sigma$ to make $V_{eff}$ negative while
$V_{eff} (f^2 )=0$. Hence, $\phi_c^2 =f^2$ is an unstable vacuum,
and this fact has been established in a selfconsistent way. The true vacuum
cannot
be reliably computed within the loop expansion. Clearly, if the bare theory
makes
sense, there is a true vacuum with $\phi_c^2$ of the order of $\Lambda$.
Moreover,
for large $\Lambda$, at energies $\mu \sim \Lambda$ the one loop RG equations
cannot apply, because if they did, our previous analysis indicates that the
theory would become unbounded. The RG equations first break down when none of
the
running couplings is large. Therefore, the breakdown can only be induced by
large
scaling violation terms (terms of order ${{\mu^2}\over{\Lambda^2}}$ --  if no
dimension
five operators are allowed). At this energy, where the breakdown of the RG
equations
first occurs, higher loop contributions to the scaling violations are still
suppressed.
Hence we are led to the conclusion that the scaling violations are dominated by
tree level
contributions when they kick in. If an acceptable bare theory that realizes
the Krive--Linde choice of couplings is found, the true vacuum expectation
value of $\phi$
will be of the order $\Lambda$, and the range in energy where scaling
violations overcome
the continuum terms in the $\beta$--functions should be calculable in
perturbation theory,
and dominated by the tree level term. The decay rate of the metastable
perturbative vacuum
can be estimated with only the help of the effective potential for small
classical fields and,
hence, is cutoff insensitive and calculable. However, what happens to the
``universe'' after the
decay, will be strongly cutoff dependent.

In summary, the effective potential calculation leads us one step beyond the
analysis
of the RG flow of the couplings, namely, that bounded Hamiltonians that
have their couplings adjusted so as to ostensibly violate the lower bound on
$\alpha$,
will end up having an unstable vacuum where the violation occurs.
There is no contradiction between the conclusions obtained from studying the
flow of
the couplings and the effective potential calculations; such a contradiction
would be
unacceptable because both approaches are within ordinary
renormalized perturbation theory and this is a self-consistent scheme. While
not every model with a
cutoff will lead to the realization of a local, but globally unstable minimum
in $V_{eff}$ I can't see any reason why some perfectly reasonable
cutoffs schemes shouldn't realize this.

If an unstable minimum is occurring near the origin in field space for some
small $g^2 (0)$,
it is likely that this minimum becomes stable for even smaller $g^2 (0)$
and then the simplest possibility is that, in the broken
phase, a first order transition will be induced when the bare Yukawa coupling
is increased through some critical value. No such transition has been reported
by any of the lattice studies that I am aware of. However, with possibly one
exception, nobody has been looking very hard for it either. The possible
exception is the work of Kuti and Shen\cite{KutiShen}: it was reported in the
Latt90 conference
proceedings that ``no sign of vacuum instability'' was observed; by this it
is meant that no double well structure in $V_{eff}$ has been found, although it
was searched for. Unfortunately, there is no paper or preprint yet available,
with more details, so my comments about their work are entirely based on two
brief talks that appeared in the conference proceedings of Lattice 89
and 90 and on a private communication by Yue Shen.
These authors have obtained results claimed to show that,
in what seems a weakly coupled system, renormalized perturbation theory fails
miserably. It is claimed that the structure of the one loop RG
equations that we discussed above precludes the occurrence of an unstable
vacuum in $V_{eff}$; I do not understand the argument, but the discussion above
would force me to disagree with any cutoff--scheme independent conclusion of
this type. The simulation method employed in this work\cite{KutiShen} is at
fixed
value of the zero mode of the scalar field; in the presence of two vacua a
finite volume system will phase separate and obscure the first order transition
if the
barrier faced by the particular algorithmical dynamics that
was chosen is not large enough.
Other simulation methods, of the more ordinary kind, should not suffer
from this deficiency, and a strong first order transition should be detectable.
But, as I said already, I think this wasn't something other people were looking
for.\footnote{From discussions with A. De I learned at the conference that the
numerical evidence against this first order phase transition, in the staggered
case, is by now stronger than I had previously thought.}
Moreover, more complicated effects take place at strong Yukawa couplings and
they may have come into play before the first order transition had a chance to
cut in.\footnote{
More specifically, a new phase, dominated by special lattice effects, cuts into
the space of couplings, possibly before a double well structure can develop;
more
discussion may be included in the contribution of I. Montvay.}
These effects are mainly related to what the doublers are doing; it seems
to me that an effective potential computation in such a system must include
also classical scalar fields carrying momenta ${{\pi}\over{2}}
[(1,1,1,1)+(\pm 1, \pm 1, \pm 1,\pm 1 )]$ because each one of these modes can
produce a
pair of light fermionic excitations. The suppression by the kinetic energy term
at tree level is likely not to be sufficient to subdue the fluctuations of the
higher modes of the scalar field.\footnote{ We are not even guaranteed that the
$g=0$
flavor identification will hold, nor can we take approximate Lorentz invariance
in the
low energy sector for granted. On a lattice the kinetic term is limited in its
ability to
suppress the higher momentum fluctuations of the scalar field. If it can be
ignored
altogether the model
is essentially a four fermi interaction and exhibits a phase structure similar
to the one
explained by I. Affleck\cite{Aff} in two dimensions. The phase structure of
various models at
strong $g$ is evidence to this occurring.}

It is also possible that one cannot restrict the analysis to just
the calculation of the effective potential and that higher orders in the
derivative expansion of the effective action (for each one of the 16 modes)
are needed. For the purpose of finding this first order transition
on the lattice it seems that Wilson fermions would be simpler than naive or
Susskind fermions with or without reduction. If one insists on having a chiral
model the safest would probably be to try to implement the Roman chiral Yukawa
program\cite{romanYuk}.

In summary, the single issue where one could expect some progress on the
lattice
is to find or exclude a first order phase transition inside the broken phase at
which
the scalar field expectation value increases by a jump when the bare Yukawa
coupling is
increased.

It would be nice of course to investigate a chiral Yukawa model with
Goldstone bosons in the
broken phase, just to be a little closer to nature. Also, since anyhow we are
not yet after real numbers here, we should try to find some sufficiently
interesting models that can be attacked analytically by non-perturbative means.
Calling up the old large $N$ workhorse again, George Bathas, a Rutgers graduate
student and myself have been looking for a model that has a soluble large
$N$ limit and incorporates as much of the dynamics of the Yukawa system as
possible.
All previous utilizations of large $N$ models in this context that I am aware
of either
used multi-component fermions and
single component (or a finite number of components) scalar fields, or simply
ignored scalar loops\cite{BaggNac}\cite{EinhGold}. This makes the scalars
classical
in the large $N$ limit and robs the models of the scalar dynamics
which could be
important. We would like
to allow the scalar field to fluctuate, but this is impossible to do in a model
in which
all fields are rank one tensors under the global symmetry group ($O(N),~
U(N),~Sp(N)$) because we cannot write an invariant Yukawa interaction. We
cannot make the
scalar field a rank two tensor because the dimensions four operator $Tr \Phi^4$
will generate
planar diagrams and we don't know how to sum these in dimension four. So we are
left with
the single option of making one of the fermions a rank two tensor and the other
fermion
and the scalar field a rank one tensor. This model is diagrammatically similar
to
the two dimensional 't Hooft model\cite{tHooft} and will be dominated by
``rainbow'' diagrams, because
it doesn't contain any selfcouplings between fields with two indices. It is
likely therefore, that in the large $N$ limit, the model will be reducible to a
finite
set of integro-differential equations for simple observables. More precisely,
the model is
defined as follows (for $O(N)$ to be specific):
$$
L= \bar \psi_L^{i\alpha} h_L(\partial^2 ) \sigma_\mu \partial_\mu
\psi_L^{i\alpha} +
\bar \xi_R^{\alpha} h_R (\partial^2 ) \bar \sigma_\mu \partial_\mu
\xi_R^{\alpha}+
$$
$$
{g\over{\sqrt{N}}} \bar \psi_L^{i,\alpha} \phi^i \xi_R^{\alpha}+
{g\over{\sqrt{N}}} \bar \xi_R^{\alpha} \phi^i \psi_L^{i,\alpha}+L_0
$$
$$
L_0 ={1\over{2}} \partial_\mu \phi^i
h(\partial^2 ) \partial_\mu \phi^i-{1\over 2} \mu^2 \phi^i \phi^i +
{{\lambda}\over{8 N}}
(\phi^i \phi^i )^2
$$
The global symmetry group is $O(N_1)\times O(N_2)$ where $i=1,2,...N_1$ and
$\alpha=1,2,...N_2$ and
$L$ and
$R$ stand for right handed and left handed fermions as usual. $N_1 N_2 = N^2$
and we take the limit
$N\rightarrow
\infty$ with ${{N_1}\over{N_2}} \equiv \rho $ held fixed and away from zero.
The functions $h_L$, $h_R$,
and $h$ implement some cutoff of the Pauli--Villars variety. Other cutoffs can
be easily incorporated;
in particular this model is hoped to be a place where one could study the Roman
approach to chiral
Yukawa theories\cite{romanYuk} outside perturbation theory. Also, if things
work out nicely,
one may have a reasonably rich toy model in which to investigate, outside
perturbation
theory, whether it is possible to regularize a theory that has global chiral
symmetries rendered non--anomalous by nontrivial intra--fermion cancellations
and with no ``extra baggage'' beyond the ordinary particle content
at weak couplings\cite{romanchiral}. The gauging of this kind of symmetries is
the problem that this
workshop is all about.

There is one potential strong coupling issue hidden in the Yukawa sector that I
would like to
mention. We plan to study it in our large $N$ model; it may be a more important
issue than
checking the validity of the perturbative evaluation of the ``vacuum
stability'' bound because it
is more obviously of a nonperturbative nature. Other large $N$ studies cannot
treat the scalar and
fermion dynamics when they are of similar strength in determining the vacuum
fluctuations.
Excitations that carry fermion number may be more complicated than a free
fermion
in an inert broken background\cite{bags}. A fermion may prefer to ``dig''
itself a ``
hole'' (or a ``bag'' of the SLAC
variety) in the condensate and loose some energy in the process (the scalar
field gains some energy, but the
net outcome is possibly still better than leaving the condensate inert).
This might imply that he Yukawa coupling measured from scalar to two fermion
decay, for example, be very different from the Yukawa coupling defined from the
ratio of the fermion mass to the stiffness ($f$) of the scalar condensate. If
the scalar particle is sufficiently heavy it may be that $m_f$ is
significantly less than the product of $f$ by the coupling extracted from
the decay of the scalar into a fermion anti-fermion pair. So it
could be that while the fermion looks safely light the
system is strongly coupled in reality with a dynamic coupling much closer to
the unitarity bound. It should be possible to almost literally ``see'' this
effect on the
lattice. Again one needs a reasonably strong Yukawa coupling.

\begin{acknowledge}

I wish to thank the organizers for giving me the opportunity to participate in
this
very pleasant workshop.
I am grateful to Urs Heller for innumerable discussions on most of the above.
Thanks are due to him and also to Pavlos Vranas and Markus Klomfass for a very
pleasant collaboration. I also benefited from some
comments by Michael Peskin on the issue of vacuum stability and from Shmuel
Nussinov on the $S$
parameter. I acknowledge a private communication by Yue Shen in response
to some questions I had about his work. This research was supported in part by
the DOE under grant nr. DE-FG05-90ER40559.

\end{acknowledge}


\begin{thebibliography}{99}
\bibitem{Einhbook}
M. B. Einhorn, editor, {\bf The Standard Model Higgs Bososn} , Vol 8 in the
``Current Physics -- Sources and Comments'' series, North Holland (1991).

\bibitem{Sherrev}
M. Sher, {\em Phys. Rev. Lett.} {\bf 45} (1980) 1131.

\bibitem{EllFog}
J. Ellis, G. L. Fogli, {\em Phys. Lett.} {\bf B249} (1990) 543.

\bibitem{EllFogLis}
J. Ellis, G. L. Fogli,  E. Lisi, {\em Phys. Lett.} {\bf B274} (1992) 456.

\bibitem{neubcapri}
H. Neuberger, {\em Nucl. Phys.} {\bf B17} (Proc. Suppl.) (1990) 17.

\bibitem{Georgibook}
H. Georgi, {\bf Weak Interactions and Modern Particle Theory}, Addison-Wesley
(1984).

\bibitem{Georgitalk}
H. Georgi, talk delivered at this conference.

\bibitem{Sylel}
K. Symanzik in {\bf ``Recent Developments in Gauge Theories''}, (Cargese 1979),
Eds. G. 't Hooft
et. al., Plenum Press, NY (1980); {\em Nucl. Phys.} {\bf B226} (1983) 187.

M. L\"uscher, in {\bf Critical Phenomena, Random Systems, Gauge Theory}  (Les
Houches 1984), Eds. K. Osterwalder, R. Stora, North-Holland, Amsterdam (1986).

G. Parisi,  {\em Nucl. Phys.} {\bf B254} (1985) 59.

\bibitem{LW3}
M. L\"uscher and P. Weisz, {\em Phys. Lett.} {\bf B212} (1988) 472; {\em Nucl.
Phys.} {\bf B318} (1989) 705.

\bibitem{HCMC}
J. Kuti, L. Lin and Y. Shen, {\em Phys. Rev. Lett.} {\bf 61} (1988) 678; A.
Hasenfratz,
K. Jansen, J. Jers\'ak, C. B. Lang, T. Neuhaus, H. Yoneyama, {\em Nucl. Phys.}
{\bf
B317} (1989) 81; G. Bhanot, K. Bitar, {\em Phys. Rev. Lett.} {\bf  61} (1988)
427.

\bibitem{BBHNnum}
G. Bhanot, K. Bitar, U. M. Heller, H. Neuberger, {\em Nucl. Phys.} {\bf B353}
(1991) 551.

\bibitem{BBHNana}
G. Bhanot, K. Bitar, U. M. Heller, H. Neuberger, {\em Nucl. Phys.} {\bf B343}
(1990) 467.

\bibitem{neubtalla}
H. Neuberger, in  {\bf ``Lattice Higgs Workshop''} eds. Berg et. al., World
Scientific (1988).

\bibitem{EinhN}
M. Einhorn, {\em Nucl. Phys.} {\bf B246} (1984) 75.

\bibitem{Dashneub}
R. Dashen, H. Neuberger, {\em Phys. Rev. Lett.} {\bf 50} (1983) 1897.

\bibitem{Lin}
 L. Lin, J. Kuti, Y. Shen, in {\bf ``Lattice Higgs Workshop''} eds. Berg et.
al., World Scientific (1988).

 \bibitem{LW12}
M. L\"uscher and P. Weisz, {\em Nucl. Phys.} {\bf B290[FS20]} (1987) 25; {\em
Nucl. Phys.} {\bf B295[FS21]} (1988) 65.

\bibitem{Kuti1}
J. Kuti, Y. Shen, {\em Phys. Rev. Lett.} {\bf 60} (1988) 85.

\bibitem{HNV}
U. Heller, H. Neuberger, P. Vranas, FSU--SCRI--91--94, (1991).

\bibitem{HKNV}
U. Heller, M. Klomfass, H. Neuberger, P. Vranas, talk at Latt91, Tsukuba,
Japan, (1991).

\bibitem{neubf4}
H. Neuberger, {\em Phys. Lett.} {\bf 199B} (1987) 536.

\bibitem{zimmer}
F. Zimmermann, talk at Latt91, Tsukuba, Japan, (1991).

\bibitem{brown}
L. Brown, R. Goble, {\em Phys. Rev. Lett.} {\bf 20} (1968) 346; {\em Phys.
Rev.} {\bf D3} (1971) 723.

\bibitem{spar}
M. Peskin, T. Takeuchi, SLAC--PUB--5618, (1991).

\bibitem{neubfpiprl}
H. Neuberger, {\em Phys. Rev. Lett.} {\bf 59} (1987)1877.

\bibitem{hellerneub}
U. Heller, H. Neuberger, {\em Phys. Lett.} {\bf 207B} (1988)189 .

\bibitem{leut}
J. Gasser, H. Leutwyler, {\em Phys. Lett.} {\bf B188} (1987) 477.

\bibitem{BitVra}
K. Bitar, P. Vranas, FSU--SCRI--92--54, (1992).

\bibitem{phen}
G. Alatarelli, CERN TH.6317/91 (1991); P. Langacker, UPR--0492T (1992).

\bibitem{SherZ}
M. Lindner, M. Sher, W. Zaglauter, {\em Phys. Lett} {\bf B228} (1989) 139.

\bibitem{revYuk}
R. Shrock, ITP--SB--91--42 (1991); A. De, J. Jers\'ak, HLRZ--91--83 (1991).

\bibitem{KrLd}
I. Krive, A. Linde,  {\em Nucl. Phys.} {\bf B117} (1976) 265.

\bibitem{Cabetal}
N. Cabibbo, L. Maiani, G. Parisi, R. Petronzio, {\em Nucl. Phys.} {\bf B158}
(1979) 295.

\bibitem{KutiShen}
Y. Shen, J. Kuti, L. Lin, P. Rossi, {\em Nucl. Phys.} {\bf B9} (Proc. Suppl.)
(1989) 99; Y. Shen, {\em Nucl. Phys.} {\bf B20} (Proc. Suppl.) (1991) 613.

\bibitem{Aff}
I. Affleck, {\em Phys. Lett.} {\bf 109B} (1982) 307.

\bibitem{romanYuk}
A. Borrelli, L. Maiani, G. C. Rossi, R. Sisto, M. Testa, {\em Phys. Lett.} {\bf
B221} (1989) 360.

\bibitem{BaggNac}
J. Bagger, S. Naculich, {\em Phys. Rev. Lett.} {\bf 67} (1991) 2252;
JHU--TIPAC--910018 (1991).

\bibitem{EinhGold}
M. Einhorn, G. Goldberg, {\em Phys. Rev. Lett.} {\bf 57} (1986) 2115.

\bibitem{tHooft}
G. 't Hooft, {\em Nucl. Phys.} {\bf B75}(1974)461.

\bibitem{romanchiral}
A. Borrelli, L. Maiani, G. C. Rossi, R. Sisto, M. Testa, {\em Nucl. Phys.}{\bf
B333} (1990)335.

\bibitem{bags}
P. Vinciarelli, {\em Lett. Nuovo Cimento} {\bf 4} (1972) 905; T. D. Lee, G. C.
Wick, {\em Phys. Rev.}  {\bf D} (1974)2291; W. Bardeen, M. Chanowitz, S. Drell,
M. Weinstein, T.-M. Yan, {\em Phys. Rev.} {\bf D10}  (1974) 1094; R. MacKenzie,
F. Wilczek, A. Zee, {\em Phys. Rev. Lett} {\bf 53} (1984) 2203; F. Wilczek,
IASSNS--HEP--90/20 (1990).






\end{thebibliography}
\end{document}